\documentclass[pre]{revtex4b3}

\usepackage{dcolumn}
\usepackage{amsmath}
\usepackage{amssymb}
\usepackage{epsfig}
\usepackage{multicol}

\DeclareMathAlphabet{\mabold}{OT1}{cmr}{bx}{it}
\newcommand\mean[1]{\ensuremath{\langle#1\rangle}}
\newcommand{\meanl}[1]{\left\langle #1 \right\rangle}
\newcommand{\kb}{\mabold{k}}
\newcommand{\qb}{\mabold{q}}
\newcommand{\rb}{\mabold{r}}
\newcommand{\bs}{\mabold{s}}
\newcommand{\bnab}{\boldsymbol{\nabla}}

\begin{document}

%

\title[Roughening and superroughening in sine-Gordon models]{Roughening
and superroughening in the ordered and random two-dimensional
sine-Gordon models}

\author{Angel S\'anchez}\email{anxo@math.uc3m.es}%
\homepage{http://valbuena.fis.ucm.es/~anxo}

\affiliation{Grupo Interdisciplinar de Sistemas Complicados (GISC),
Departamento de Matem\'aticas,\\ Universidad Carlos III de Madrid, 
Avda.\ Universidad 30, 28911 Legan\'es, Madrid, Spain}   

\author{A.\ R.\ Bishop}

\affiliation{Theoretical Division and Center for Nonlinear Studies,
MS B258,
Los Alamos National Laboratory, Los Alamos, NM 87545, U.S.A.}

\author{Esteban Moro}

\affiliation{Theoretical Physics Department, University of Oxford,
1 Keble Road, OX1 3NP, United Kingdom}
           
\date{\today}

\begin{abstract}
We present a comparative numerical study of the 
ordered and the random two-dimensional sine-Gordon models on a lattice.
We analytically compute the main features
of the expected high temperature phase of both models, described
by the Edwards-Wilkinson equation.
We then use those results to locate the transition 
temperatures of both models in our Langevin dynamics simulations.
We show that our results reconcile previous contradictory numerical
works concerning the superroughening transition in the random sine-Gordon
model.
We also find evidence supporting the existence of two different 
low temperature phases for the disordered model. 
We discuss our results
in view of the different analytical predictions available and 
comment on the nature of these two putative phases. 
\end{abstract}


\maketitle

\begin{multicols}{2} 

\section{Introduction} 

The location and characteristics of phase transitions in 
disordered media constitute a long-standing and 
controversial question, particularly in
more than one spatial dimension \cite{Plischke}.
The question becomes even more difficult if the disorder is not very weak;
then, new, non-trivial behavior is commonly found,
involving features such as aging, ergodicity breaking, extremely
slow dynamics, complicated energy landscapes, etc.; major examples
of this are spin glasses and structural glasses \cite{Plischke,Young}. 
In this context, 
the properties of crystal surfaces growing on disordered
substrates, frequently
described by a two-dimensional
random-phase sine-Gordon model (RsGM), 
have attracted a lot of attention in the last decade
\cite{Toner,Tsai,Batrouni,Riegercom,Cule,Cule2,Krug,Scheidl,%
Marinari,Lancaster,Zeng,%
Tsai2,Rieger1,Coluzzi,us1,jjfises,Rieger2,Raul,Hwanew,Rieger3}.
Interestingly, the same
model describes many other relevant physical problems, such as 
randomly pinned 
flux lines confined in a plane \cite{Batrouni,Hwanew,Fisher,Hwa2,Blatter},
vortex
lines in planar Josephson junctions \cite{Vinokur}, charge density 
waves \cite{Fukuyama}, and commensurate-incommensurate transitions
\cite{Villain}.

In spite of those efforts, the 
phase diagram and main features of the RsGM are not clear yet.
To summarize what is known, 
we refer for comparison 
to the {\em ordered} sine-Gordon model (OsGM), which is rather
well understood (see, e.g., 
\cite{Plischke,Weeks,Beijeren,Barabasi,Krug2,Villain2} and
references therein).
The hamiltonian for the OsGM and the RsGM is
\begin{equation}\label{h}
\begin{split}
{H} = \sum_{\rb}\Big[&\frac{1}{2} \sum_{n.n.}[h(\rb)-h(\rb')]^2 +\\
&+V_0\Big(1-\cos[h(\rb)-h^{(0)}(\rb)]\Big)\Big].
\end{split}
\end{equation}
where $n.n.$ stands for the nearest neighbors of site $\rb$.
The OsGM corresponds to 
$h^{(0)}(\rb)\equiv 0$, whereas the RsGM
is defined by choosing the {\em quenched} disorder variables
$h^{(0)}(\rb)$ randomly from a uniform
distribution in $[0,2\pi]$. We will discuss our work in
terms of growth on flat (OsGM) or disordered (RsGM) 
substrates (see \cite{Hwa2,Blatter,Vinokur,Fukuyama,Villain} for other
physical interpretations): Accordingly, $h(\rb)$ is 
a continuous variable representing
the height of the growing surface at site $\rb$ of the lattice, and 
the cosine term is a potential energy making integer 
(i.e., multiples of the crystalline lattice constant) values of the height
energetically favorable. We will 
consider two-dimensional (2D) square lattices, i.e., $\rb=(r_1,r_2)$,
with $N=L\times L$ sites.
As first shown by Chui and Weeks \cite{Chui} (see also \cite{Knops})
by means of a renormalization group (RG) approach \cite{Golden}, the
OsGM posseses a Kosterlitz-Thouless \cite{KT} type 
topological transition between 
a low temperature, flat phase and a high temperature, rough phase,
the latter being described by the Edwards-Wilkinson (EW) equation 
\cite{EW}, i.e., the diffusion equation with 
additive white noise (see below). Above the
so called roughening temperature ($T_R$), thermal fluctuations 
effectively suppress the effect of the cosine potential, and the 
surface becomes free, described only by the kinetic part of
the hamiltonian (\ref{h}). As will be shown below, 
the most important measurable consequence
of this is that the width of the interface, 
\begin{equation}
\label{guiz}
w^2=\frac{1}{N}\sum_{\rb}(h(\rb)-\bar{h})^2,\quad \bar{h}=
\frac{1}{N} \sum_{\rb} h(\rb),
\end{equation}
scales (in 2D) as $w^2\sim\ln L$ in the asymptotic regime. 

In contrast with the clear picture for the OsGM, 
there are very 
few generally accepted results for the RsGM. One 
of them is that there must be a roughening 
temperature above which the potential effectively vanishes (much as
in the case of the OsGM) leading to EW behavior. 
Apart from this, theoretical predictions about the low temperature 
phase largely disagree (a good summary is given in the third paper
in Ref.\ \cite{Cule}): While RG calculations predict a {\em superrough}
low temperature phase, with $w^2\sim \ln^2L$, replica-symmetry
breaking variational approaches lead to $w^2 \sim \ln L$ independently
of temperature.
Numerical simulations were not very conclusive either: 
Batrouni and Hwa \cite{Batrouni} did not find evidence for an equilibrium
phase transition in Langevin dynamics, although Monte 
Carlo simulations by 
Rieger \cite{Riegercom} showed a transition from 
a superrough phase to a EW phase 
for stronger potentials [larger
$V_0$ in Eq.\ (\ref{h})] than those used by 
Batrouni and Hwa. Subsequent numerical work
\cite{Marinari,Lancaster,Coluzzi,jjfises} presented more evidence
of superrough ($\ln^2L$) behavior, albeit with
large quantitative discrepancies with the predictions of RG theories.
Finally, a number of works using special optimization algorithms 
\cite{Zeng,Rieger1,Rieger2} or direct numerical simulations \cite{us1}
strongly supported superrough behavior at zero temperature. Very 
recently \cite{Hwanew}, simulations of a related model provided
more evidence of $\ln^2$ behavior at finite temperatures, although
this model did not allow study of the transition. In summary, 
most researchers believe that there is 
indeed a superrough low temperature phase in the RsGM, but its nature
(glassy or not), the transition temperature, and its dependence on 
the model parameters remain unclear. 

In this paper, we attempt to shed light on these issues
by simultaneously studying the OsGM and the RsGM
in different regions of their phase diagram. As we will show
below, it turns out that the potential strength, $V_0$,
crucially determines the model features. In addition, it is 
also natural to ask about 
the {\em intensity of the 
disorder}: How does the model phenomenology change if the disorder
takes values in $[0,\epsilon]$ with $\epsilon<2\pi$? 
The importance of these points can be clearly seen in
the Langevin equation, 
\begin{equation}
\label{lange}
\frac{dh(\rb,t)}{dt}=-\frac{\delta H}{\delta h(\rb,t)}+\eta(\rb,t),
\end{equation}
where $\eta(\rb,t)$ is a gaussian white noise of zero mean and correlations
obeying the fluctuation-dissipation theorem,
\begin{equation}
\label{ruido}
\mean{\eta(\rb',t')
\eta(\rb,t)} = 2 T \delta^{(2)}(\rb-\rb') \delta(t-t'),
\end{equation}
where
$T$ stands for the temperature of 
the system, $\langle\ldots\rangle$ indicates thermal averages (over $\eta$)
and 
Boltzmann's constant is set to
$k_B=1$.
For our hamiltonian, Eq.\ (\ref{h}), the Langevin equation is
\begin{equation}\label{uno}
\begin{split}
\frac{\partial h(\rb,t)}{\partial t} = & \nabla^2 h(\rb,t) -\\
&- V_0 \sin[h(\rb,t)-h^{(0)}(\rb)] + \eta(\rb,t),
\end{split}
\end{equation}
and changing 
variables according to 
$\tilde{h}(\rb,t) = h(\rb,t) - h^{(0)}(\rb)$ (i.e., the height referred to 
the substrate), we find
\begin{equation}\label{dos}
\begin{split}
\frac{\partial \tilde h(\rb,t)}{\partial t} = &\epsilon \tilde F(\rb)+
 \nabla^2 \tilde h(\rb,t) -\\
&- V_0 \sin(\tilde h(\rb,t)) + \eta(\rb,t)
\end{split}
\end{equation}
where $F(r) = \nabla^2 h^{(0)}(\rb)/\epsilon$. In this form, 
the disorder is a random (correlated in space)
chemical potential acting on a surface growing on 
a flat substrate.  If we think of the roughening transition for the OsGM
in terms of the interplay between the temperature $T$ and the
energy scale introduced by the potential, $V_0$, when $\epsilon \neq 0$,
we have another energy scale, $\epsilon$, which can modify the 
universal features of the roughening transition or even give rise to
novel thermodynamical transitions. 

Having 
the above issues in mind, we discuss our work according to the 
following scheme: Section II presents an analytical study of 
the EW equation and other
statistical mechanics results about the energy and roughness. 
Section III deals with the main part of our 
work, namely Langevin dynamics simulations of the OsGM and 
the RsGM, beginning with $V_0=1$ and $\epsilon=2\pi$
and subsequently
analyzing the model behavior for different $V_0$ and
$\epsilon$.
Finally, in 
Sec.\ IV we present and discuss our conclusions and 
indicate future directions.

\section{Analytical results} 

\subsection{Linear theory: The Edwards-Wilkinson equation}

According to the RG
approach \cite{Weeks}, the high temperature phase of the 
OsGM 
obeys the EW equation \cite{EW},
\begin{equation}\label{eqs-apen}
\frac{\partial h(\rb,t)}{\partial t} =  {\bnab}^2
h(\rb,t) + \eta(\rb,t).
\end{equation}                              
Equation (\ref{eqs-apen}) can be 
solved by means of the Fourier decomposition (see \cite{Krug2}
for details),
\begin{equation}\label{modosFourier}
\hat h_{\qb} = \frac{1}{L} \sum_{\rb}
\mathrm{e}^{i \qb \cdot \rb} h({\rb},t),
\end{equation}                                                               
where 
$\qb = \frac{2\pi}{L} \kb$,
$k_i = 0,\ldots,L-1$ is the reciprocal vector. The structure factor 
can then be shown to be 
\begin{equation}\label{sdek}
S(\qb) = \mean{\hat h_{\qb}\hat h_{-\qb}} = T \frac{1-\mathrm{e}^{-2  \omega_{\qb} t}}{\omega_{\qb}},
\end{equation}                                        
$\omega_{\qb}$ being the 2D EW discrete dispersion 
relation
\begin{equation}\label{omegaEW2D}
\omega_{\qb}=4 \sin^2\left(\frac{q_1}{2} \right)
 +4\sin^2\left(\frac{q_2}{2} \right).
\end{equation}
{}From $S(\qb)$ we can obtain the relevant magnitudes,
such as the total roughness
\begin{equation}\label{w2apen}
w^2(t) = \meanl{\frac{1}{L^d}\sum_{\rb}[h(\rb,t)-\bar h]^2} = \frac{1}{L^d} \sum_{\qb\neq 0} S(\qb),
\end{equation}                                
the 
correlation function,
\begin{eqnarray}\label{crapen}       
C(\rb) &=& \meanl{\frac{1}{L^2}\sum_{\bs}[h(\bs+\rb)-h(\bs)]^2} \nonumber
\\ &=& \frac{2}{L^2} \sum_{\qb} S(\qb) [1-\cos(\qb\cdot\rb)].
\end{eqnarray}                                  
the total slope, 
\begin{equation}\label{s2apen}
s^2(t) = \frac{1}{L^2} \sum_{\qb\neq 0} S(\qb) \omega_{\qb},
\end{equation}
and the slope difference correlation function, 
\begin{equation}\label{grapen}
\begin{split}
G(\rb) &= \meanl{\frac{1}{L^2}\sum_{\bs}[\bnab h(\bs+\rb)-\bnab h(\bs)]^2} =\\
&= \frac{4}{L^2}\sum_{\qb} S(\qb) \sum_{i=1}^2 [1-\cos(q_i r_i)][1-\cos(q_i)].
\end{split}
\end{equation}

{}From Eq.\ (\ref{sdek}) we can find an estimate for 
the time needed to reach saturation, $t_{\times}$,
and the dynamic exponent, $z$: We compute the time that the structure
factor needs to be within a 1\% of its saturated form for the 
slowest Fourier mode, that with the lowest $|\qb|$, $|\qb|=2\pi/L$:
\begin{equation}\label{tcross2d}
t_{\times} \simeq 3 \times 10^{-2} L^2,
\end{equation}
implying 
$z = 2$. For the saturated roughness, we obtain
\begin{equation}\label{w2EW2D}
w^2(t \to \infty,L)=
\frac{T}{4 L^2}\sum_{k_1,k_2=1}^{L-1}
\left[\sin^2\left(\frac{q_1}{2} \right)+\sin^2\left(\frac{q_2}{2} \right)
\right]^{-1},   
\end{equation}                             
which cannot be computed exactly but, for large $L$, can 
be 
approximated changing the sum by an integral and the sine functions by their 
arguments, arriving at 
\begin{eqnarray}\label{w2EW2Dapprox}
w^2(t\rightarrow\infty,L)&\simeq&\frac{1}{(2\pi)^2}
\int_{2\pi/L}^{\pi} \int_{2\pi/L}^{\pi} dq_x dq_y \frac{T}{(q_x^2+q_y^2)} 
\simeq\nonumber\\ &\simeq&\frac{T}{2\pi} \int _{2\pi/L}^{\pi} \frac{dq}{q}
 =  \frac{T}{2\pi} \ln L,
\end{eqnarray}                     
yielding a roughness exponent 
$\alpha = 0$. 
As for the total slope, it tends to a value independent of 
$L$,
\begin{equation}\label{s2EW2Dbeta}
s^2 \simeq T,
\end{equation}
whereas for the correlation functions, we find for large $r$
\begin{eqnarray}\label{crEW2D}
C(\mathbf{r})
&\simeq&  \frac{T}{ \pi} \ln r,\\
\label{grEW2D}
G(\mathbf{r})
&\simeq & 2 s^2 + \frac{T}{4\pi }\ln 
\frac{r^8}{[(r^2+1)^2-4r_1^2][(r^2+1)^2-4r_2^2]} \nonumber\\
 &\simeq &{2T}.
\end{eqnarray}            

\subsection{Other results for the energy and the roughness}

At equilibrium, the partition function for the OsGM is 
\begin{equation}\label{funpar}
{Z} = \int \bigg[\prod_{\rb} dh(\rb)\bigg]\ e^{-\beta [\frac12 \sum_{\rb}
 (\nabla h(\rb))^2 + V_0(1 -\cos h(\rb))]},
\end{equation}
where $\beta = T^{-1}$.
Expanding for high temperatures \cite{berez} means 
rewriting Eq.\ (\ref{funpar}) as a series in powers of $\beta V_0$:
\begin{eqnarray}\label{funpar_exp}
{Z} &=& \sum_{n=0}^{\infty}\frac{(\beta V_0)^n}{n!}
\int \bigg[\prod_{\rb} dh(\rb)\bigg]\ \mathrm{e}^{-\beta
\sum_{\rb} \frac12 (\nabla h(\rb))^2}\times \nonumber \\
&&\phantom{xxxxxxxxxxxxxx}\times\bigg(\sum_{\rb} \cos h(\rb)\bigg)^n=
\nonumber \\
\label{funpar_exp_promedios}
&=&{Z}_0 \sum_{n=0}^{\infty}\frac{(\beta V_0)^n}{n!}  
\langle \big(\sum_{\rb}\cos h(\rb)\big)^n \rangle_{{H}_0},
\end{eqnarray}                                     
where ${Z}_0$ is the partition function of $H_0$, the free hamiltonian [i.e., 
Eq.\ (\ref{h}) without the potential term].
By means of this expansion we obtain
\begin{equation}\label{promedio_cos_exp_H0}
\begin{split}
\mean{\cos h(\rb')}_{{H}}  =& \frac{{Z}_0}{{Z}}\sum_{n=0}^{\infty}\frac{(\beta 
V_0)^n}{n!}\times\\
&\times\mean{\big(\sum_{\rb} \cos  h(\rb)\big)^n \cos  h(\rb')}_{{H}_0},
\end{split}
\end{equation}                      
where subdominant terms such as  $\beta^n \exp(-A/\beta)$ have been 
neglected, and only terms of the order $\beta^n$ have been kept. 
The expression above can be put in the form 
\begin{equation}
\begin{split}
\mean{\cos  h(\rb')}_{{H}}& = \left(
{\displaystyle \sum_{n=1}^{\infty} 
\frac{2^{2n-1} (n!)^2}{n(\beta V_0)^{2n-1}}}\right)\times\\[2mm]
&\times
\left({\displaystyle
\sum_{n=0}^{\infty}\frac{(\beta V_0 )^{2n}}{2^{2n} (n!)^2} }
\right)^{-1} 
\label{promedio_cos_sumas}
= \\[2mm] = \frac{\beta V_0}{2}&-\frac{(\beta V_0)^3}{16}+
\frac{(\beta V_0)^5}{96}-
+{O}((\beta V_0)^7),
\end{split}
\end{equation} 
yielding for the approximate energy per site
\begin{equation}\label{energia_T_altas}
\begin{split}
E = &\frac{1}{L^2}\mean{{H}} = \frac{1}{2 \beta} +\\
&+ V_0
\left(1 - \frac{\beta V_0}{2}+
\frac{(\beta V_0)^3}{16} + {O}((\beta V_0)^5) \right).
\end{split}
\end{equation}     

For the roughness, we have to compute 
\begin{equation}\label{exprug1}
\begin{split}
\mean{(h(\rb)-\bar h)^2}_{H}& = \frac{Z_0}{Z}\times\\
\times \sum_{n=0}^{\infty}
\frac{(\beta V_0 )^n}{n!}&
\meanl{h^2(\rb) \sum_{\rb'} \left[\cos  h(\rb')\right]^n}_{H_0},
\end{split}
\end{equation}
assuming that at equilibrium $\bar h = 0$.
Neglecting again subdominant terms, we find 
\begin{equation}\label{exprug2}
\mean{h^2(\rb)}_{H} = \frac{Z_0}{Z}\mean{h^2(\rb)}_{H_0},
\end{equation}
and hence
\begin{equation}\label{exprug3}
\begin{split}
w^2_{H} = w^2_{H_0}\big( 1-&\frac{ \beta V_0}{4} + \frac{3 (\beta V_0)^2}{64} -\\ 
-&\frac{19 (\beta V_0)^3}{2304} + O((\beta V_0)^4) \big).
\end{split}
\end{equation}                                                

Finally, at low temperatures the height exhibits only small deviations
from $h=0$, and therefore we can approximate
the hamiltonian as 
\begin{equation}\label{HamLT}
\begin{split}
{H} &= \sum_{i}\frac{1}{2} (\nabla h(\rb))^2 + V_0[1-\cos(h(\rb))]\\
&\simeq \sum_{i}\frac{1}{2} (\nabla h(\rb))^2 +  V_0\frac{h(\rb)^2}{2},
\end{split}
\end{equation}
i.e.,
$2N$ quadratic terms, each one of which, according to the equipartition theorem
\cite{Huang},
contributes with 
$T/2$ to the energy value. Taking into account the global 
factor $1/2$, we conclude that at low temperatures
the energy of the OsGM is approximately $E=T/2$. 

\section{Langevin dynamics results}

\subsection{Numerical simulation details}

We have
integrated the Langevin equation (\ref{uno}) corresponding to the hamiltonian
(\ref{h}), on 
$L\times L$ square lattices with periodic boundary
conditions, using
a stochastic second order Runge-Kutta method \cite{simu}; 
in some cases, we have repeated the simulations with a
Heun method \cite{maxi},
with excellent agreement between both procedures. We are
therefore sure that our results are not an artifact of our numerical method,
a conclusion further reinforced by the agreement with the theoretical 
expectations of the previous section as we discuss below. The simulations
reported in this paper were carried out with a time step $\Delta t=0.01$ on
lattices of sizes $L=64, 128$ and 256.
It is important to stress that we did not perform averages over 
the quenched noise in the RsGM; however, we checked that the
outcome of our simulations did not depend strongly on the realization of
the quenched noise or the initial conditions (flat, $h(\rb)=0$; as the 
substrate, $h(\rb)=h^{(0)}(\rb)$, or random) by repeating several times a 
number of our simulations. In all cases, simulations consisted of an
equilibration time and a measuring period. Eq.\ (\ref{tcross2d}) predicts
that the time needed for equilibration is $t_{\times}\simeq 500$ for 
$L=128$ and $t_{\times}\simeq 2 000$ for $L=256$, and hence we used 
equilibration times of $5 000$ and $10 000$ units, respectively; 
afterwards, we let the system evolve for an equal period, over which 
we performed thermal averages. Equilibration was ensured in all cases
by verifying that the fluctuations of the energy were gaussian and by
checking the equality of the specific heat computed from those fluctuations
and from derivatives of the mean energy \cite{Plischke},
as well as by monitoring the evolution in time of the quantities
of interest toward a stationary state.
As an additional test, we compared
the imposed simulation temperature, arising from the noise term, to that
measured during the evolution according to the equipartition theorem 
\cite{Huang}; both quantities were always found to agree within a 
0.1\%. Finally, we did a few very longer runs, whose outcome agreed 
with that of the shorter runs.

\subsection{Standard RsGM: $V_0=1$ and $\epsilon=2\pi$}

We begin by discussing our results for the ``canonical'' version of both
the OsGM and the RsGM, i.e., hamiltonian (\ref{h}) with $V_0=1$, as
studied (for the ordered case) in \cite{falo,yo1}. 
In those works, the roughening temperature was determined 
by a direct comparison to RG predictions, looking
for the temperature at which the height difference correlation function
reached a universal (in the RG framework) value, with the result that
for the OsGM $T_R\approx 25$ in our dimensionless units. 
Remarkably, this is the RG value for $T_R$, which makes 
very tempting to claim that this method indeed yields $T_R$ 
correctly. Monte Carlo simulations 
of the discrete gaussian model (hamiltonian (\ref{h}) 
with $V_0=0$ and $h(\rb)$ restricted to integer values)
by Shugard {\em et al.}\ \cite{Shugard} with the same criterion
for locating the transition yielded similar results.
However, as RG calculations are perturbative
in $V_0$ and carried out on the continuum Langevin
equation \cite{Weeks,Beijeren,Barabasi,Villain2,Chui,Knops},
it is not obvious that they apply to a discrete model with $V_0=1$,
i.e., of the same order as the kinetic term. In view of this, we decided
to include the OsGM in this study, both to analyze in detail whether the
comparison to the universal RG prediction for the factor is a good tool
to find $T_R$ and to compare its high 
temperature phase 
with that of the RsGM, which should also be of EW type.

\begin{figure}
\begin{center}
\epsfig{file=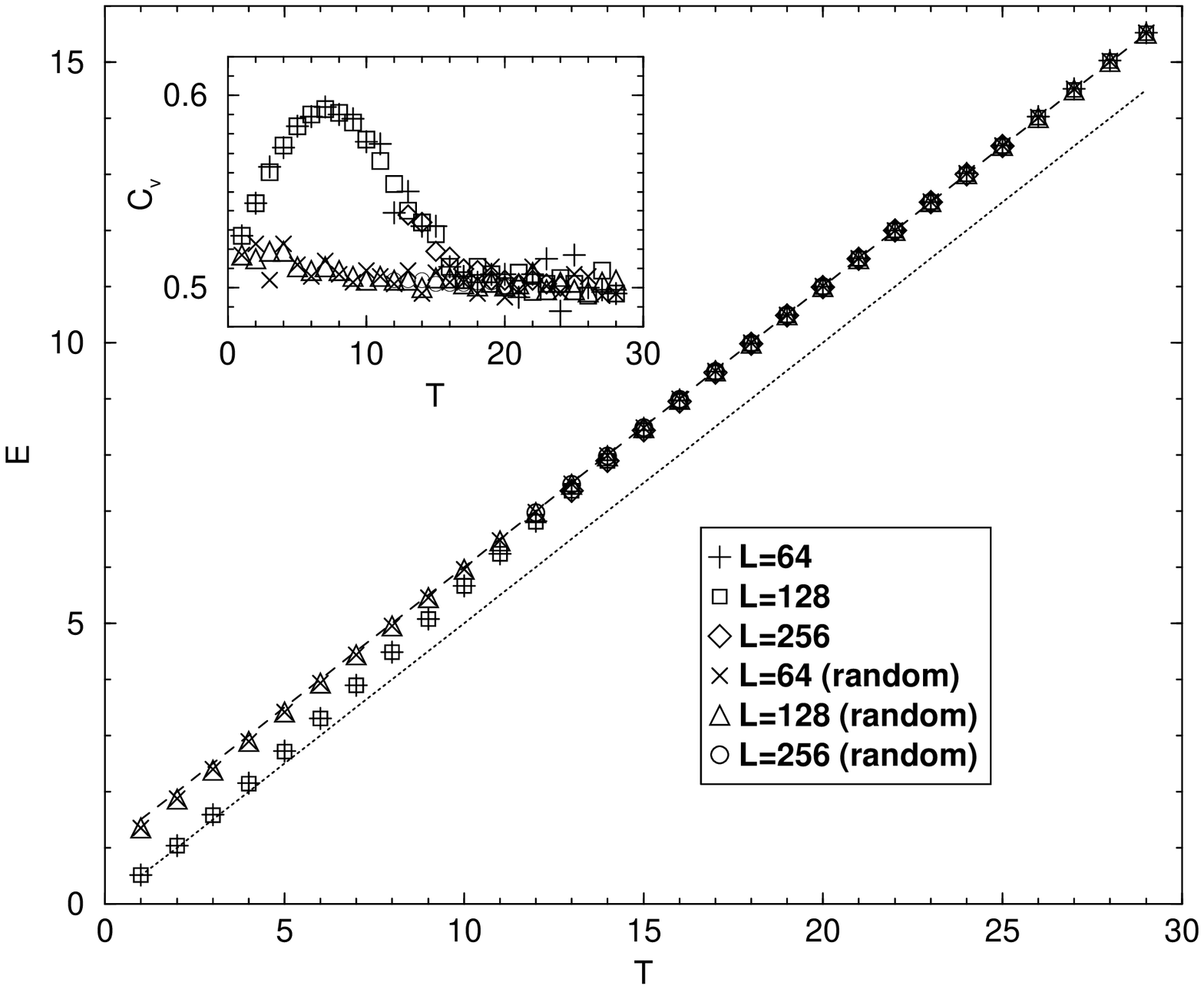, width=2.7in, angle=-90}
\caption{Mean energy for the OsGM and the RsGM
vs temperature.
Symbols as indicated in the plot. The straight lines
correspond to the low temperature prediction (lower line) and the high 
temperature prediction (upper line). Inset: Specific heat vs temperature;
symbols as in the main plot. Error bars are smaller than the symbol
size.}
\label{fig2}
\end{center}
\end{figure}

The first quantity we discuss, 
shown in Fig.\ \ref{fig2}, is the mean energy of both models. As we 
see, the results are largely independent of the system 
size, and hence it is unlikely that they are affected by finite size 
effects. The plot shows that the mean energy of the
OsGM reaches the high temperature approximation at
$T_0=16\pm 1$; on the other hand, the mean energy for the RsGM is never
too far from it, although for temperatures lower than $T_1=4\pm 1$ the 
numerical values lie slightly below the high temperature result. At 
temperatures higher than $T_0$ the energies of both models 
coincide within the accuracy of our simulations. These results 
suggest that $T_0$ could be the roughening temperature, $T_R$,
for the OsGM and
$T_1$ the superroughening temperature, $T_{SR}$,
for the RsGM, because the EW behavior of 
the mean energy of both models 
indicates the effective suppresion of the sine term by temperature.
The inset in Fig.\ 
\ref{fig2} presents 
the specific heat, $C_v$, of both models, exhibiting a well
defined peak in $C_v$ for the OsGM with its maximum at temperature 
$T=9$, much lower than $T_0$. In contrast, 
we do not observe any peak for the RsGM; there might be
a peak at $T=3$, but the evidence is not conclusive. 
Concerning the peak for the OsGM, we stress again the absence
of any finite size effect, consistent with a KT type transition.
Furthermore, we believe that this peak is a (Schottky) {\em anomaly}
\cite{Golden}
similar to that observed in 2D XY and easy-plane Heisenberg spin models
\cite{Kawabata}
{\em above} the KT transition. 
Recall that when mapping the OsGM to the XY model the temperature of 
the former maps to the inverse temperature of the latter \cite{Chui}, hence
the observation of the anomaly below the possible transition temperature
$T_0$. This reinforces our interpretation of $T_0$ as the roughening 
temperature of the OsGM. 

\begin{figure}
\begin{center}
\epsfig{file=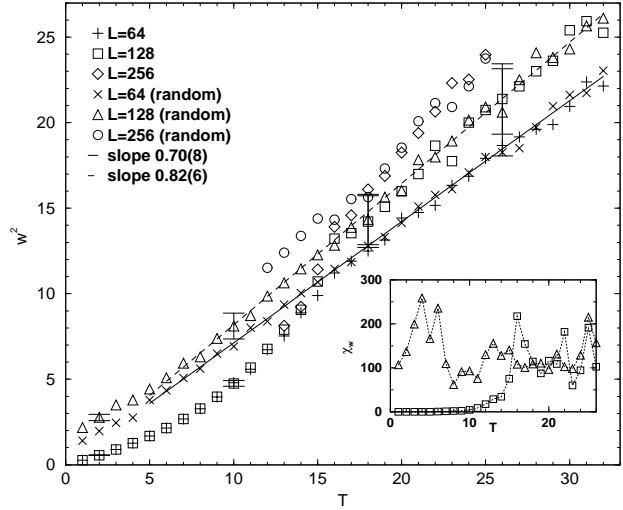, width=2.7in, angle=-90}
\caption{Total roughness for the OsGM and the RsGM
vs temperature.
Symbols as indicated in the plot. The straight lines
correspond to the high 
temperature prediction for the two sizes considered.
Inset: Roughness susceptibility vs temperature;
symbols as in the main plot. Examples for error bars in different
regions are shown.}
\label{fig3}
\end{center}
\end{figure}

The total roughness of both models, 
shown in Fig.\ \ref{fig3}, 
behaves similarly for temperatures higher
than $T_0$, depending linearly on temperature. In both cases, we 
see that the slope of the roughness depends on the system size, as 
predicted by Eq.\ (\ref{w2EW2D}). Whereas the approximation in
Eq.\ (\ref{w2EW2Dapprox})
expression yields slopes
0.66 and 0.77 for $L=64$ and 128, respectively,
if we numerically compute the exact result, Eq.\ (\ref{w2apen}), the slopes
turn out to be 0.71 y 0.82, in excellent agreement with the results 
of our simulations, 0.71 and 0.83.
Below $T_0$, the roughness for the OsGM is independent
of the size and depends nonlinearly on temperature, whereas above
$T_0$ we find linear dependence on temperature and clear finite 
size effects. For the RsGM, the linear behavior extends all the way down
to $T_1$, and below $T_1$ the behavior becomes nonlinear. The slope of 
the linear region is approximately
the same in the first part, from $T_1$ to $T_0$, and in the second, above
$T_0$, i.e., the whole linear region is well described by the EW model. 
This means that above
$T_0$ the linear model describes accurately the behavior of the 
OsGM, and hence from now on we identify $T_0$ with the roughening 
temperature $T_R$, whereas 
for the RsGM, the same is true of
$T_1$ and $T_{SR}$. 
{}Figure \ref{fig3} presents also results for 
the roughness susceptibility, $\chi_{w}$, defined 
as $\chi^2_{w}=[\langle (w^2)^2\rangle-\langle w^2
\rangle^2]/T^2$. For the OsGM, $\chi_{w}$ exhibits a very clear 
peak at $T_R$, and above $T_R$ it is the same as for
the RsGM; however, this magnitude
is very noisy and these results must be taken with 
caution. In fact, one could identify a peak for the
RsGM at $T_{SR}$, but different realizations lead to different results,
in contrast with the peak for the OsGM, which is the same for all 
realizations. Below $T_R$, the values of $\chi_w$ for the OsGM are 
independent of the system size, whereas above $T_R$ they increase 
with size without 
any definite scaling. 

{}Figure \ref{fig4} depicts the height-difference 
correlation function for the two studied models,
and shows that above $T_R=T_0$ and $T_{SR}=T_1$
they behave as predicted
by the linear
theory: The slope of the numerical height-difference correlation
function is $0.32$, indistinguishable
from the predicted $0.318$ by Eq.\ (\ref{crEW2D}).
In addition, the
correlation functions for the OsGM and the RsGM coincide, as shown 
in the plot for the RsGM. We see that the behavior of the 
correlation functions is in full agreement with our claims 
regarding $T_R$ and $T_{SR}$, and this is further confirmed by 
the plot of the slope correlation function in Fig.\ \ref{fig6}.
It is important to note that, below $T_{SR}$, the behavior of
the height difference correlation function for the RsGM is approximately 
a squared logarithm, as predicted by RG calculations. We postpone 
discussion of this point to the next section.

Finally
we studied another magnitude, namely
\begin{equation}
\label{m1}
m_1=\mean{\cos \left[h(\rb)-h^{(0)}(\rb)\right]}.
\end{equation}
For the OsGM this is the average
computed in the preceding section, whereas for the RsGM it 
is the average of the cosine of the height referred to the substrate.
Figure \ref{fig8} shows our results:
The high temperature approximation,
Eq.\ (\ref{promedio_cos_sumas}),
agrees very accurately 
with the simulations for temperatures above $T_R$ (OsGM) and 
$T_{SR}$ (RsGM). The results for both models are again indistinguishable
for temperatures above $T_R$. Interestingly,
$m_1$ is largely independent not only of the system
size, but also on the realization of the quenched disorder for the 
RsGM. 

\subsection{Other potential strengths}

We now turn to the question of the influence of the 
potential strength
on the RsGM behavior. We have
considered two representative values, 
$V_0=0.2$ and $V_0=5$, i.e., five times smaller and larger, respectively,
than the ``canonical'' value $V_0=1$. The smallest value is 
close to that considered in \cite{Batrouni}, 
$V_0=0.15$, and we expect that our results will be comparable
to theirs. 
As before, we begin by discussing the 
total roughness and the 
specific heat (see Fig.\ \ref{fig14}). First of all, for all values
of $V_0$ there is a temperature above which the roughness value is 
independent of $V_0$ and of the presence or absence of disorder. This
means that our identification of this regime with the effective 
suppresion of any potential effect is indeed correct: Different 
$V_0$ leads only to different transition temperatures. Thus, for the 
OsGM we find $T^{V_0=0.2}_R=13\pm 1$ and $T^{V_0=5}_R=19\pm 1$, in agreement
with the intuitive expectation that larger potentials need higher 
temperatures to be suppressed. Aside from this, the general shape of
the roughness curve is basically the same for the three values of 
$V_0$. The situation (for the OsGM) is the same as far as the specific
heat is concerned: Larger (smaller) $V_0$ leads to larger (smaller)
anomalies, which are displaced to higher (lower) $T$ following the 
corresponding $T_R$. 
Therefore, we conclude that changing $V_0$ 
does not introduce anything qualitatively new in the OsGM. 

The picture for the RsGM is substantially different:
Modifying $V_0$ does give rise to qualitatively 
new phenomena. 
Let us first look at the small $V_0$ case. Figure
\ref{fig14} shows that the roughness follows a straight line all 
the way down to $T=0$ (although we cannot exclude that there are 
nonlinear effects for $T\lesssim 1$ with our present resolution). This 
would suggest that there is no transition in this case, very much 
like the results of Batrouni and Hwa \cite{Batrouni}. The upper panel of 
Fig.\ \ref{fig13}, where the height difference correlation function
is depicted, confirms this interpretation, showing no dependence on
temperature in the analyzed range; Fig.\ \ref{fig12}, for 
$m_1$, agrees with this as well, in so far as the 
dependence of $m_1$ on temperature is well described by the high 
temperature expansion. In view of this, we can conclude that if
there is a transition, it occurs at a temperature smaller than $T\simeq 1$.

Finally, let us consider the large $V_0$ case, $V_0=5$. The plot of
the roughness in Fig.\ \ref{fig14} exhibits a striking
peak for $T=5$, after which the roughness
decreases until reaching the high temperature regime (marked by the 
corresponding OsGM with the same value for $V_0$) at $T=8\pm 1$. 
To assess the relevance of this peak in the roughness, we repeated our
simulations for $L=64$ and performed additional ones for $L=128$. The 
results, collected in Fig.\ \ref{fig14bis}, show that the peak is a 
realization dependent feature. However, in this plot we also see that
for temperatures below $T=5$ the roughness is roughly independent of
the system size, something which we did not observe when $V_0=1$ (the
lines for $V_0=1$ are included in Fig.\ \ref{fig14bis} again for 
comparison). Hence, even if the peak at $T=5$ does not actually exist,
that temperature does seem to separate two different regions. 
In addition,
for $V_0=5$ the specific heat has a (more smeared) maximum 
at about the same temperature as that of the roughness, although our 
data are much noisier and we cannot establish clearly the maximum 
temperature; the dependence, however, is manifestly non-monotonic. 
{}Figure \ref{fig12} supports our conclusion 
that $T_{SR}^{V_0=5}=8\pm 1$, whereas nothing special is seen as $m_1$ 
goes through $T=5$, the roughness maximum. The most intriguing 
result is the one in Fig.\ \ref{fig13} for the height difference 
correlation function: For $T\leq 5$, the scaled correlations decrease with 
temperature but, simultaneously, the correlation length increases,
up to $T=5$, when it increases beyond the system size. Above that 
temperature, it follows the same evolution as the $V_0=1$ case, finally
reaching $T_{SR}^5=8\pm 1$. Whereas in this intermediate temperature 
regime the height difference correlation functions are well described
by squared logarithms, Fig.\ \ref{fig13} immediately shows that the lowest
temperature correlations can by no means be considered squared logarithms. 
This suggest the presence of a new phase transition at $T^*=5\pm 1$. 
We will discuss the possible nature of the low temperature phase and
the existence of this transition in the next section. 

\subsection{Other disorder strengths}

We now generalize the RsGM and let
$h^{(0)}(\rb)$ to be randomly chosen from a uniform
distribution in $[0,\epsilon]$, with $0<\epsilon<2\pi$; $\epsilon=2\pi$
is the case studied in the preceding Section. As representative examples
we have considered $\epsilon=0.8$ and $\epsilon=0.2$, closer to the 
RsGM and the OsGM, respectively. 
{}From the roughness dependence on
temperature, shown in Fig.\ \ref{fig9}, we see that
for both values of $\epsilon$ the dependence
of the roughness on temperature is qualitatively similar to the OsGM,
the results for the lower $\epsilon$ value being practically the same
as for the $\epsilon=0$ case. However, the case $\epsilon=0.8$ is 
somewhat different: The low temperature region appears to consist of
two straight lines, changing slope at a temperature around $T'=5$, 
rather than a nonlinear dependence. 
By reasoning as above, we identify $T_{SR}^{\epsilon=0.2}\simeq T_R=16\pm 1$ 
and $T_{SR}^{\epsilon=0.8}=12\pm 1$, values which are confirmed by the 
energy behavior (not shown), the height difference (Fig.\ \ref{fig10}) 
and slope (not shown) correlation functions, and by the dependence of
$m_1$ on temperature (not shown).

An interesting question arises from Fig.\ \ref{fig10}: 
There is no evidence 
about the squared logarithmic behavior found for the RsGM and,
furthermore, the plots exhibit a finite correlation length below $T_{SR}$
for both values of $\epsilon$. Another intriguing fact is the nonmonotonic
dependence of the correlation function on temperature for $\epsilon=0.8$: 
{}From the curve for $T=1$, the scaled correlation function decreases 
up to $T=5$;
upon further increasing the temperature, the evolution of the curves is 
very similar to that of the OsGM. This might be connected with the change
in slope in the roughness curve mentioned in the preceding paragraph (see
Fig.\ \ref{fig9}), but we have not been able to draw a clearer connection.
All this is clear evidence
that the behavior of the RsGM is significantly dependent on 
the disorder strength.

\section{Discussion and conclusions} 

Let us begin the discussion of the above results by 
analyzing our findings about the OsGM. 
Our simulations strongly support that
$T_R=16\pm 1$
for the OsGM on a lattice with $V_0=1$.
This is in contrast
to the claims in \cite{falo,yo1} that $T_R=25$ 
where a different way of definining the transition, which assumes the
validity of the RG approach, was used (see
Sec.\ III B and \cite{falo,yo1,Shugard}).
Further, the result is also in contrast with the RG prediction itself
\cite{Weeks,Beijeren,Barabasi,Villain2,Chui},
which in our units is $T_R=8\pi$. However, we believe that the 
comparison with the linear theory for the EW high temperature phase 
has a much more physical character while keeping the basic RG ideas, and
establishes beyond doubt that for the studied lattices the roughening
temperature is $T_R=16\pm 1$ for $V_0=1$.
Another hint in favor of our claim is the finding of the 
(Schottky) anomaly in the
specific heat, which should appear below the transition temperature in 
view of what occurs for the XY and related models \cite{Kawabata}. 
Finally, the fact that we obtain the same results for both the OsGM and 
the RsGM above $T_R$ is clear evidence that the potential is 
irrelevant (in the RG sense) in that regime and that we have indeed located
the transition.
Clearly,
we cannot exclude the possibility that 
working on even larger lattices we would find the transition where
the RG predicts it, but 
the absence of
any finite size effects even for $L=256$ makes this possibility quite 
unlikely. Another possible reason for the discrepancy is the fact that our
simulations are intrinsically discrete in space while RG theories for the 
OsGM are always applied to the continuum equation; again, very much larger
lattices would remove this objection and clarify the 
effects of discreteness. Aside from that,
we have also found
that increasing (decreasing) $V_0$ increases (decreases) the roughening 
temperature: In the cases we studied, we found $T_R^{V_0=0.2}=13\pm 1$ and
$T_R^{V_0=5}=19\pm 1$, which is intuitively reasonable as
larger potential barriers require larger temperatures for the surface to 
overcome them. On the other hand, RG calculations 
are perturbative in $V_0$, so one would expect better agreement to the 
RG prediction for $V_0=0.2$, but in fact the agreement is worse in that 
case. 

Let us now turn to
the RsGM. In the ``canonical'' case, $V_0=1$, we found  a superroughening
transition at $T_{SR}=
4\pm 1=T_R/4$, to be compared to RG predictions that it
should occur at $T_R/2$.
Below $T_{SR}$,
we have obtained a $\ln^2$ dependence
of the height difference correlation function, in agreement with RG
results.
However, we have clearly shown
that the superroughening transition temperature depends
on $V_0$, confirming the earlier report by Batrouni and Hwa \cite{Batrouni}
on the absence of the transition in Langevin dynamics simulations 
for $V_0=0.15$, and later reports by Rieger \cite{Riegercom} and
 Ruiz-Lorenzo
\cite{jjfises} who also observed this dependence in Monte Carlo simulations.
In the opposite case, $V_0=5$, we find that $T_{SR}^{V_0=5}=8\pm 1$, 
considerably higher than the $V_0=1$ temperature. 
This disagrees with the RG predictions of a universal $T_{SR}$
independent of $V_0$. We believe that the agreement between our results
and the previous ones \cite{Batrouni,Riegercom,jjfises} indeed supports a
dependence of $T_{SR}$ on $V_0$, whose explanation remains an open question
as far as RG is concerned.

\begin{figure}
\begin{center}
\epsfig{file=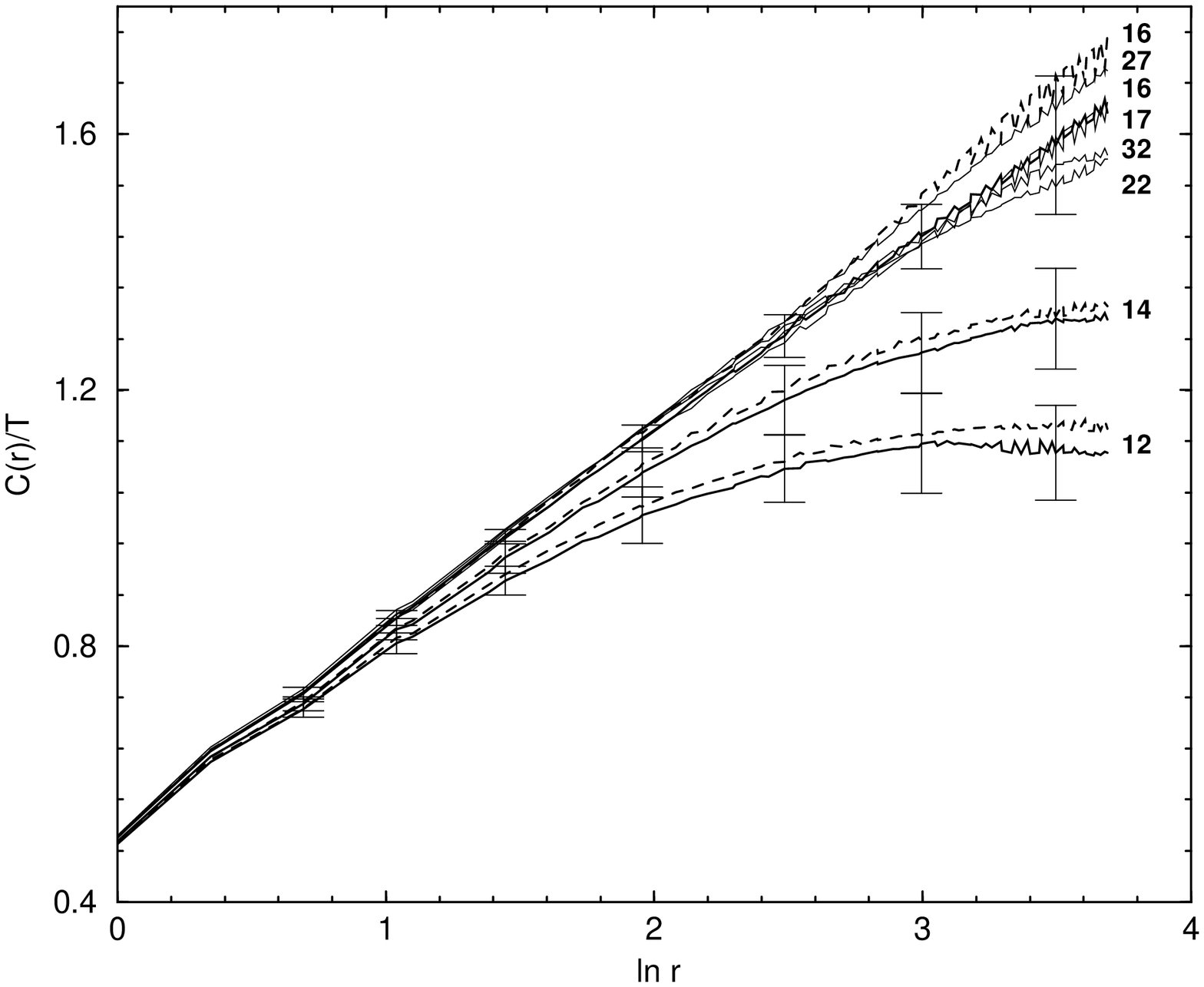, width=2.7in, angle=-90}
\epsfig{file=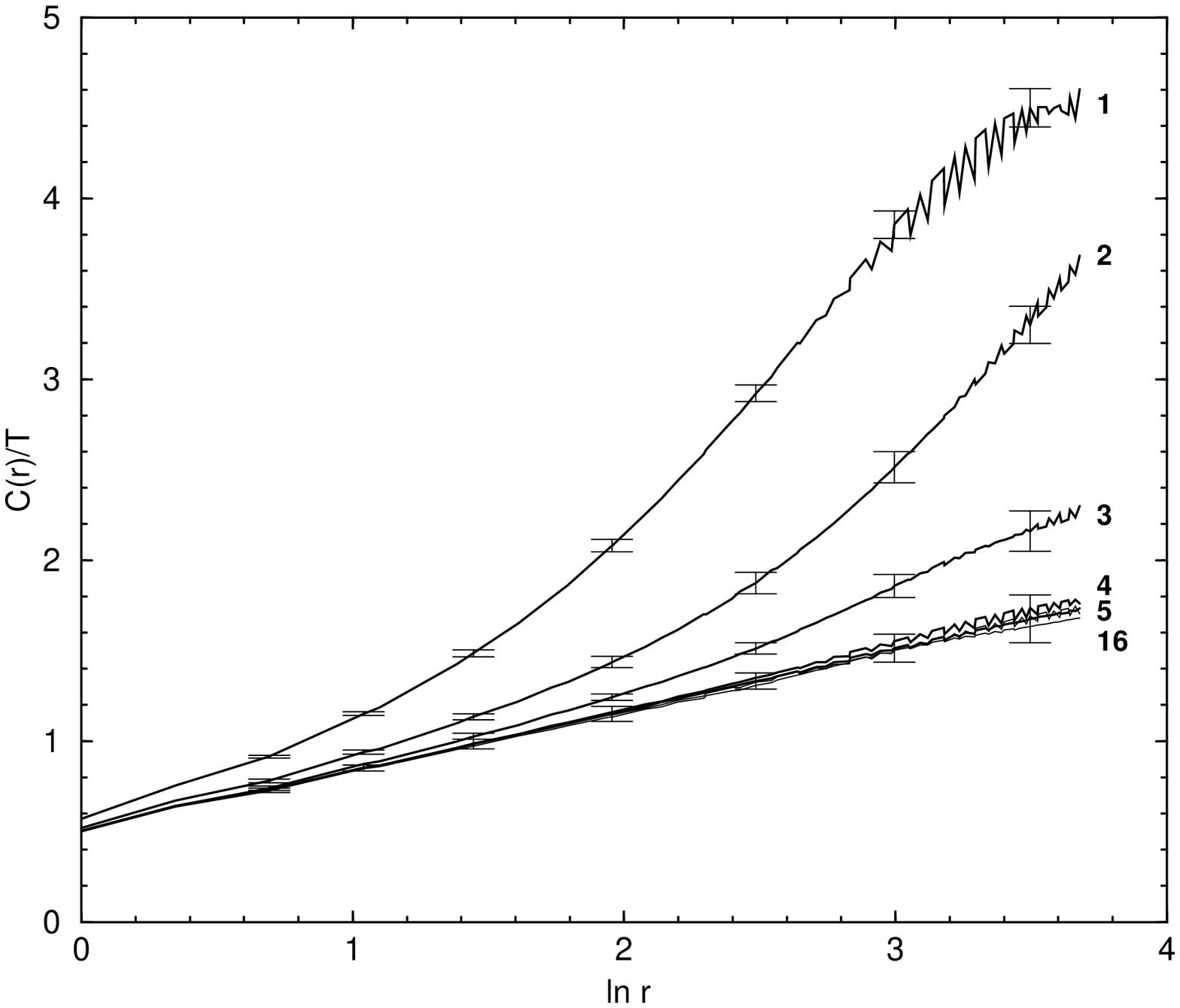, width=2.7in, angle=-90}
\caption{Height difference correlation functions (scaled by $T$) 
vs $\ln r$, $r$ being distance,
for the OsGM (upper panel) and disordered (lower 
panel) sine-Gordon model. Temperatures are indicated at the right 
side of the plots. The curve marked with temperature $16$ in the plot
for the RsGM is the correlation function for the OsGM at that temperature,
showing clearly that both overlap.}
\label{fig4}
\end{center}
\end{figure}

\begin{figure}
\begin{center}
\epsfig{file=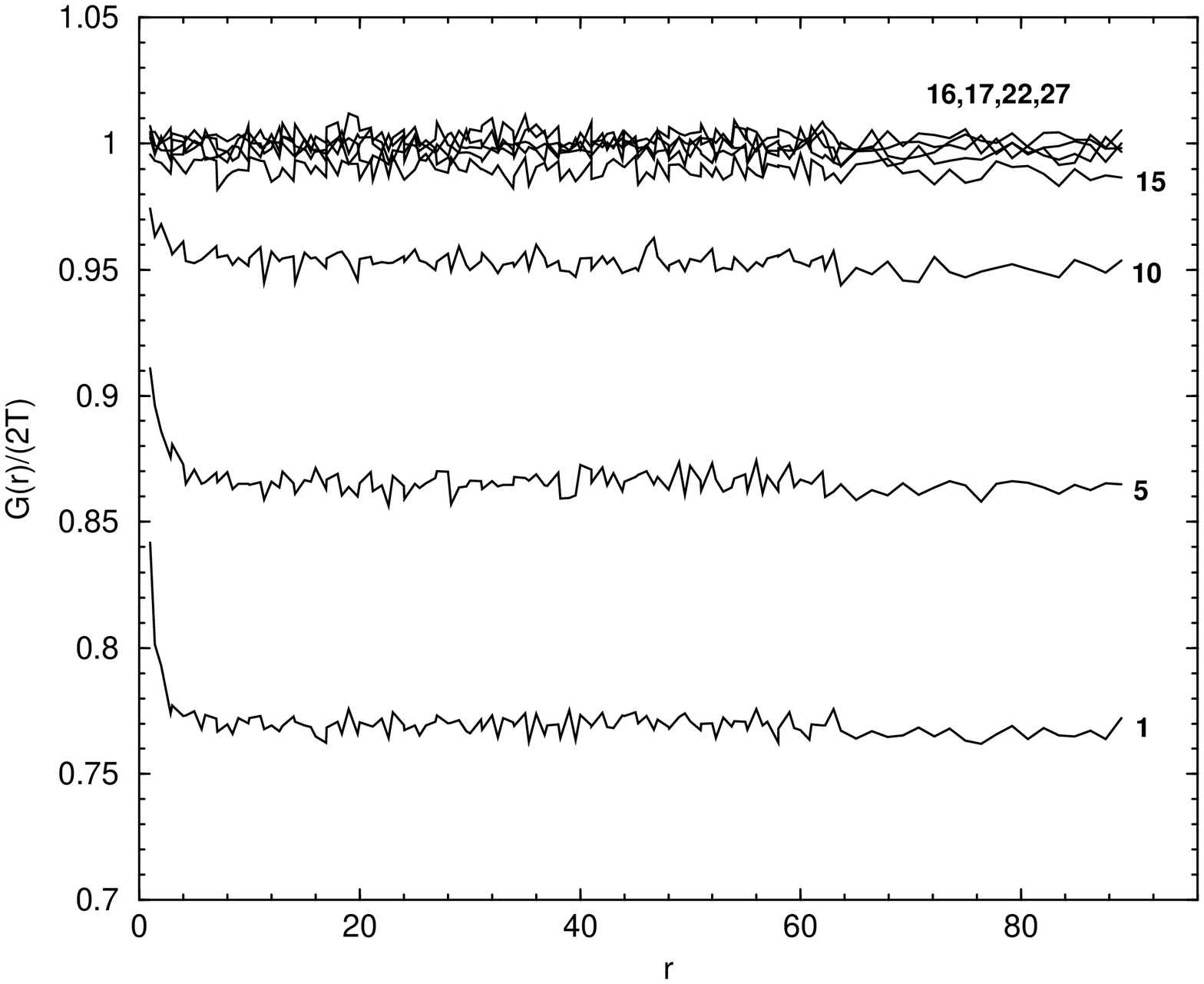, width=2.6in, angle=-90}
\epsfig{file=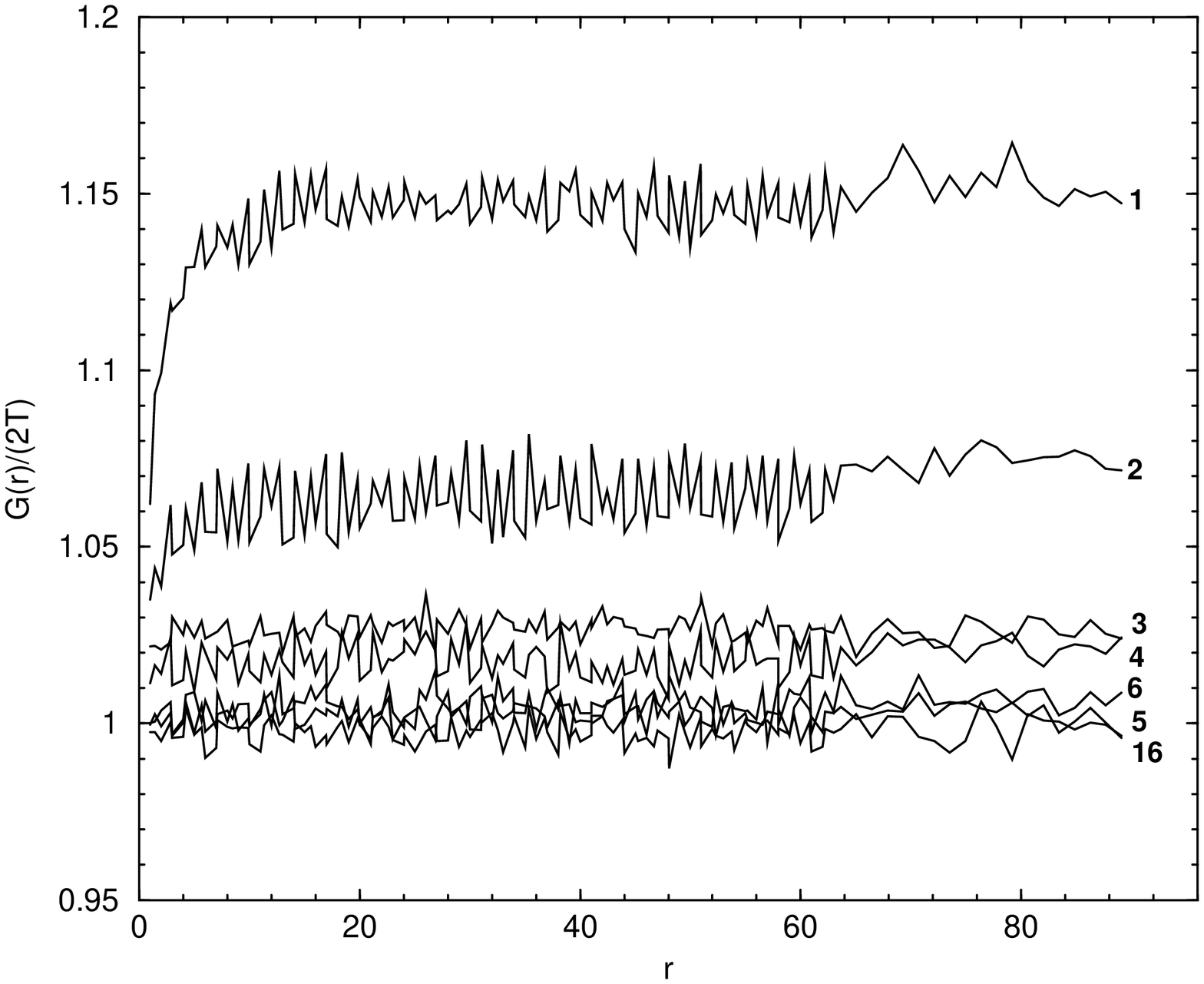, width=2.7in, angle=-90}
\caption{Slope correlation functions (scaled by $T$)
vs $\ln r$, $r$ being the distance,
for the OsGM (upper panel) and disordered (lower
panel) sine-Gordon model. Temperatures are indicated at the right
side of the plots. The curve marked with temperature $16$ in the plot
for the RsGM is the correlation function for the OsGM at that temperature,
showing clearly that both overlap.}
\label{fig6}
\end{center}
\end{figure}

\begin{figure}
\begin{center}
\epsfig{file=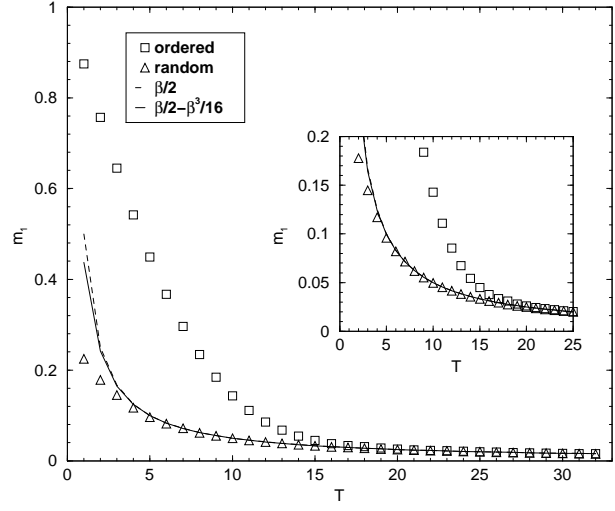, width=2.7in, angle=-90}
\caption{Comparison of the averages of the cosine term, $m_1$, 
of the OsGM and the RsGM at 
different temperatures. Symbols and lines as indicated in the 
plot; lines correspond to theoretical approximations up to order 
$\beta$ and $\beta^3$,
whereas symbols are numerical results.
Results are independent of the system size and
of the realization of the disorder; error bars for thermal averages are
smaller than the symbol size. The inset is an enlargement of the low
$m_1$ region of the plot.} 
\label{fig8}
\end{center}
\end{figure}

\begin{figure}
\begin{center}
\epsfig{file=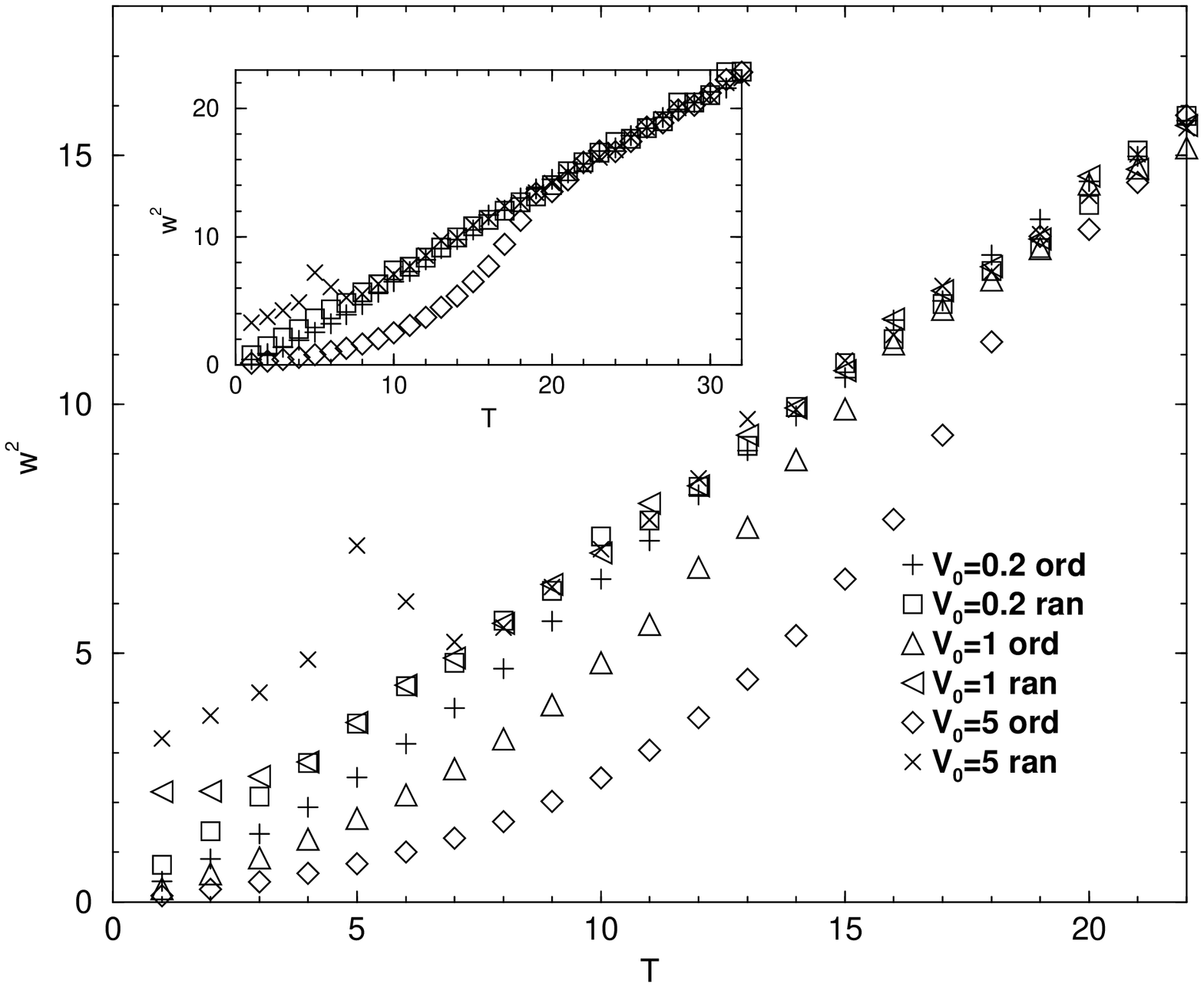, width=2.7in, angle=-90}
\epsfig{file=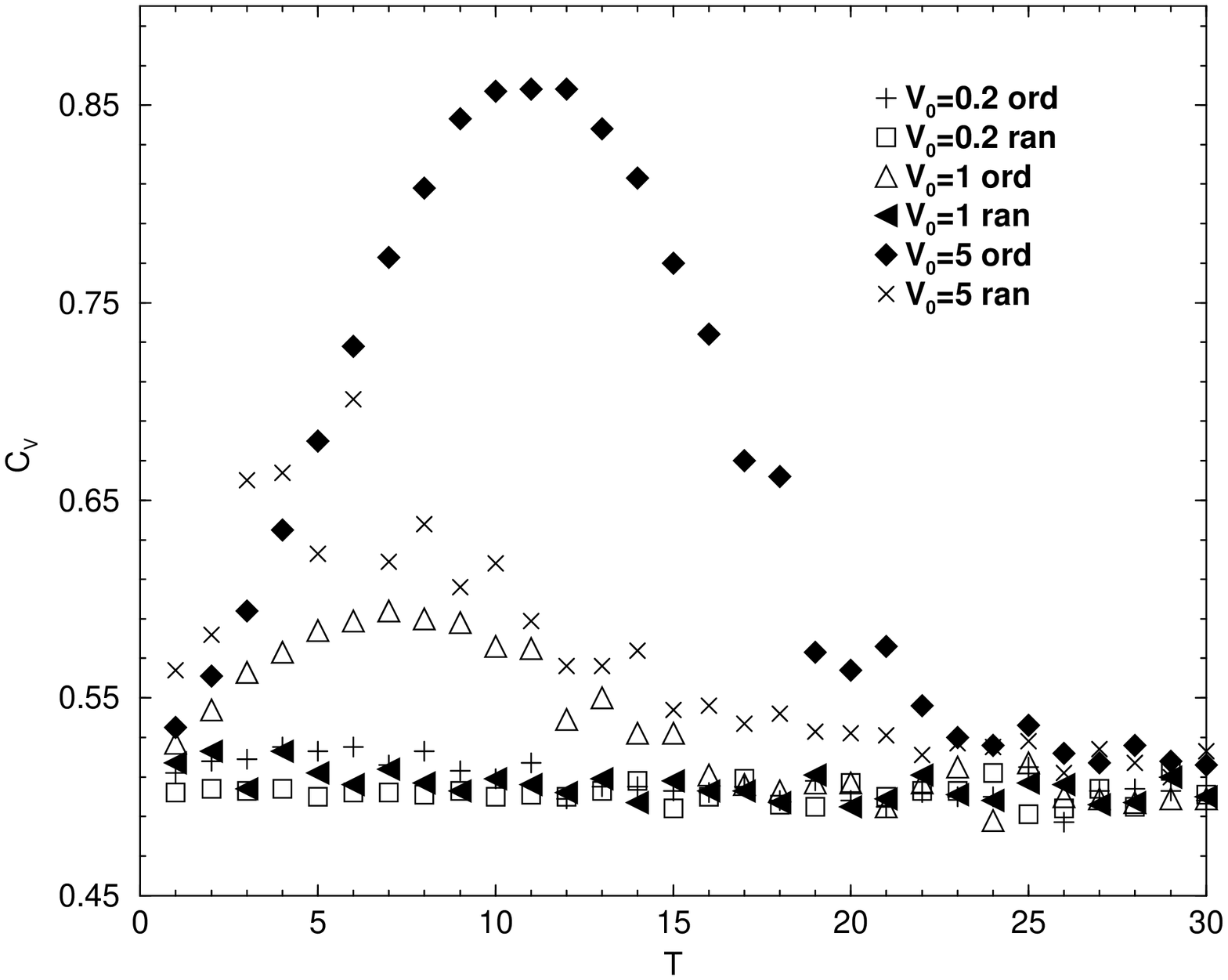, width=2.7in, angle=-90}
\caption{Comparison of the roughness (upper panel) 
and specific heat (lower panel) for the 
OsGM and the RsGM 
for different values of $V_0$.
The inset in the upper panel shows a larger range of temperatures
for the roughness dependence. Symbols as indicated in the 
plot. All the results have been obtained for $L=64$. 
Note especially the peak in the specific heat for the RsGM with $V_0=5$.}
\label{fig14}
\end{center}
\end{figure}

\begin{figure}
\begin{center}
\epsfig{file=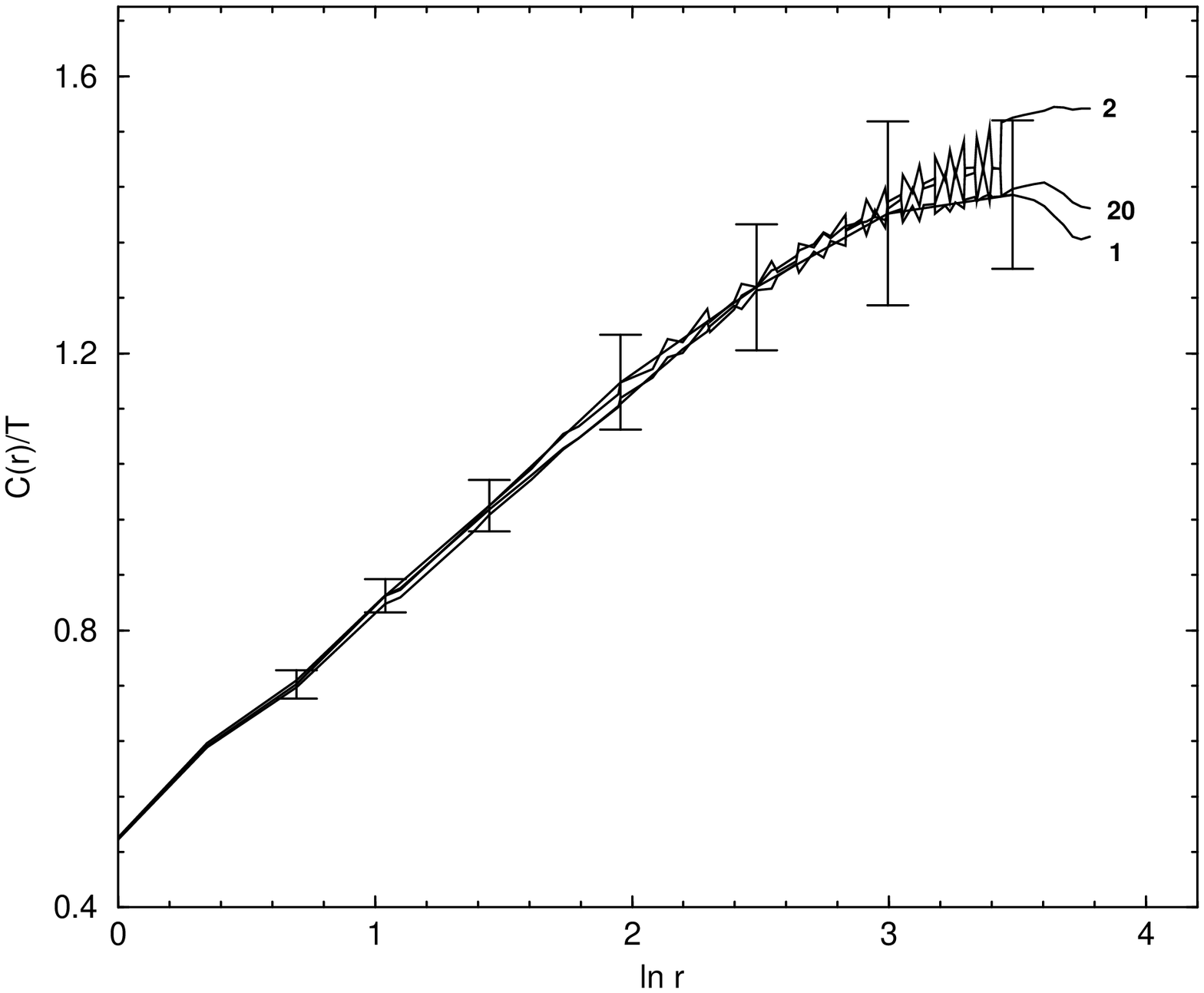, width=2.7in, angle=-90}
\epsfig{file=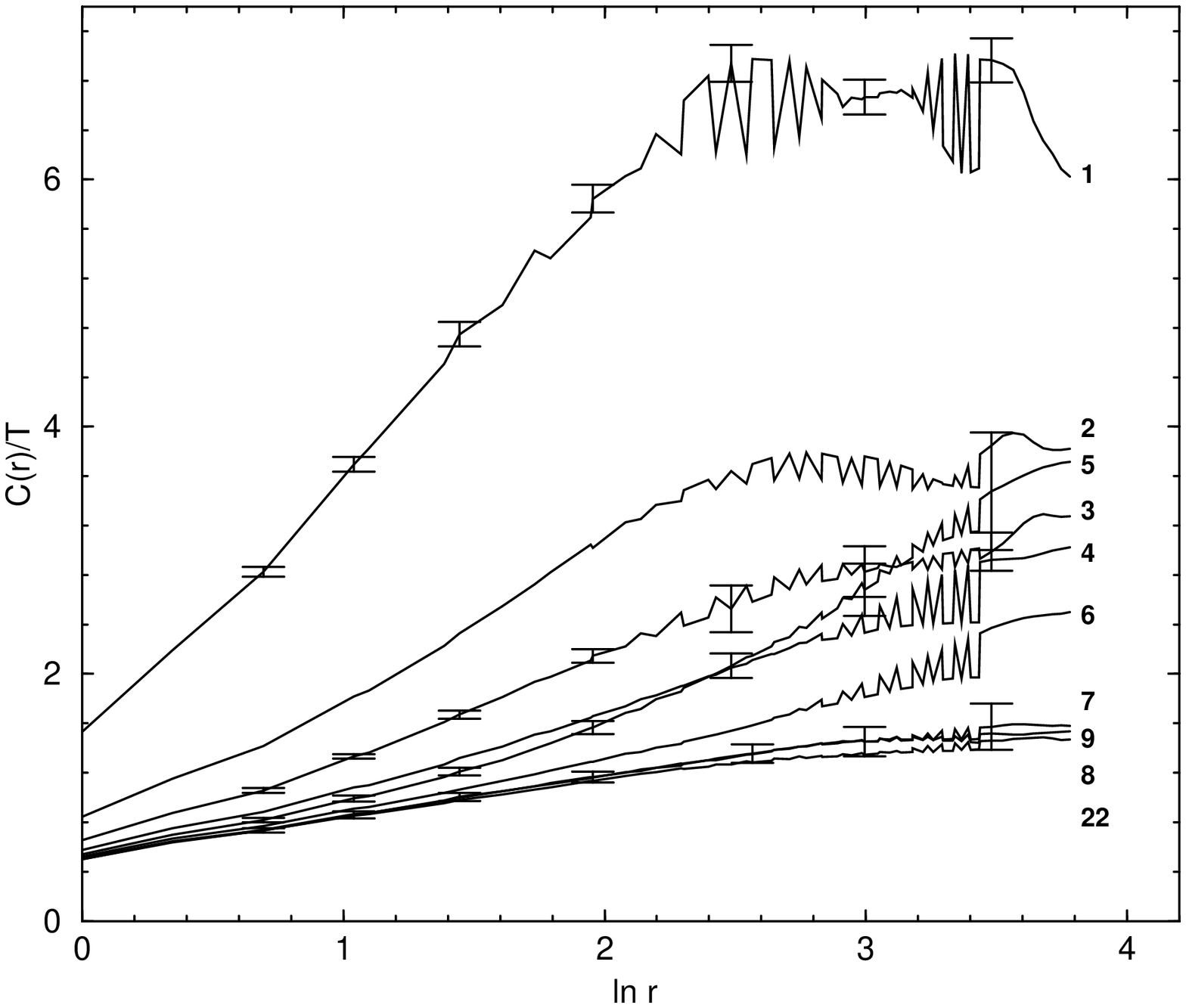, width=2.7in, angle=-90}
\caption{Comparison of the height difference correlation functions
for the RsGM with $V_0$, $V_0=0.2$ (upper panel) and 
$V_0=5$ (lower panel). 
Temperatures as indicated in the plots.}
\label{fig13}
\end{center}
\end{figure}

\begin{figure}
\begin{center}
\epsfig{file=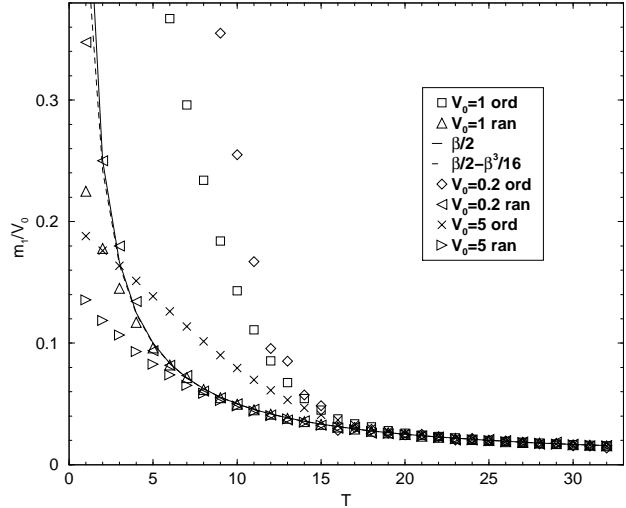, width=2.7in, angle=-90}
\caption{Comparison of the averages of the cosine term, $m_1$, 
of the OsGM and the RsGM 
for different values of $V_0$. Symbols as indicated in the plot,
and lines correspond to the theoretical approximations for high
temperature.}
\label{fig12}
\end{center}
\end{figure}

\begin{figure}
\begin{center}
\epsfig{file=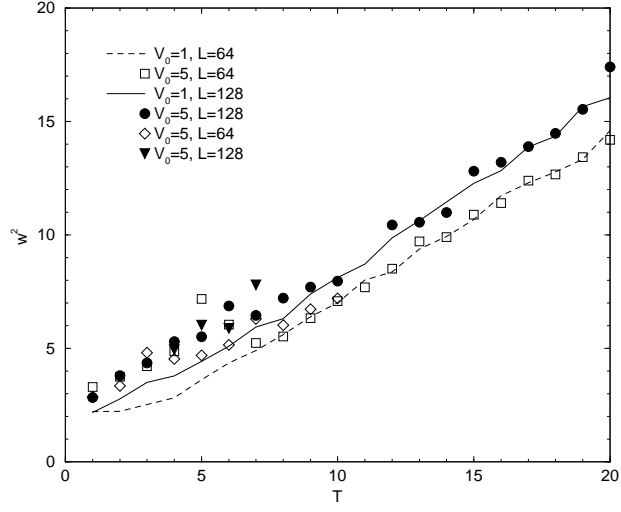, width=2.7in, angle=-90}
\caption{Comparison of the roughness 
OsGM and the RsGM 
for $V_0=1, 5$ and sizes $L=64, 128$.
Symbols as indicated in the 
plot.}
\label{fig14bis}
\end{center}
\end{figure}

\begin{figure}
\begin{center}
\epsfig{file=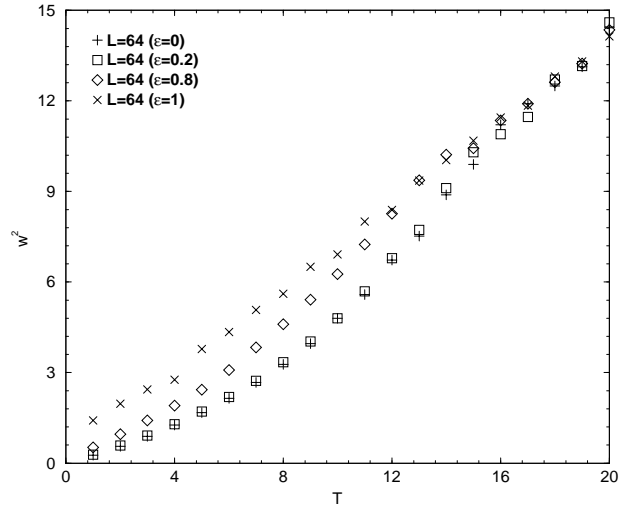, width=2.7in, angle=-90}
\caption{Comparison of the roughness for the OsGM, the RsGM,
and the two intermediate versions of the RsGM
with disorder strength $\epsilon=0.2$ and $\epsilon=0.8$ (see text for 
the corresponding definition). The size is $L=64$ in all four cases, 
and symbols are as indicated in the plot.}
\label{fig9}
\end{center}
\end{figure}

\begin{figure}
\begin{center}
\epsfig{file=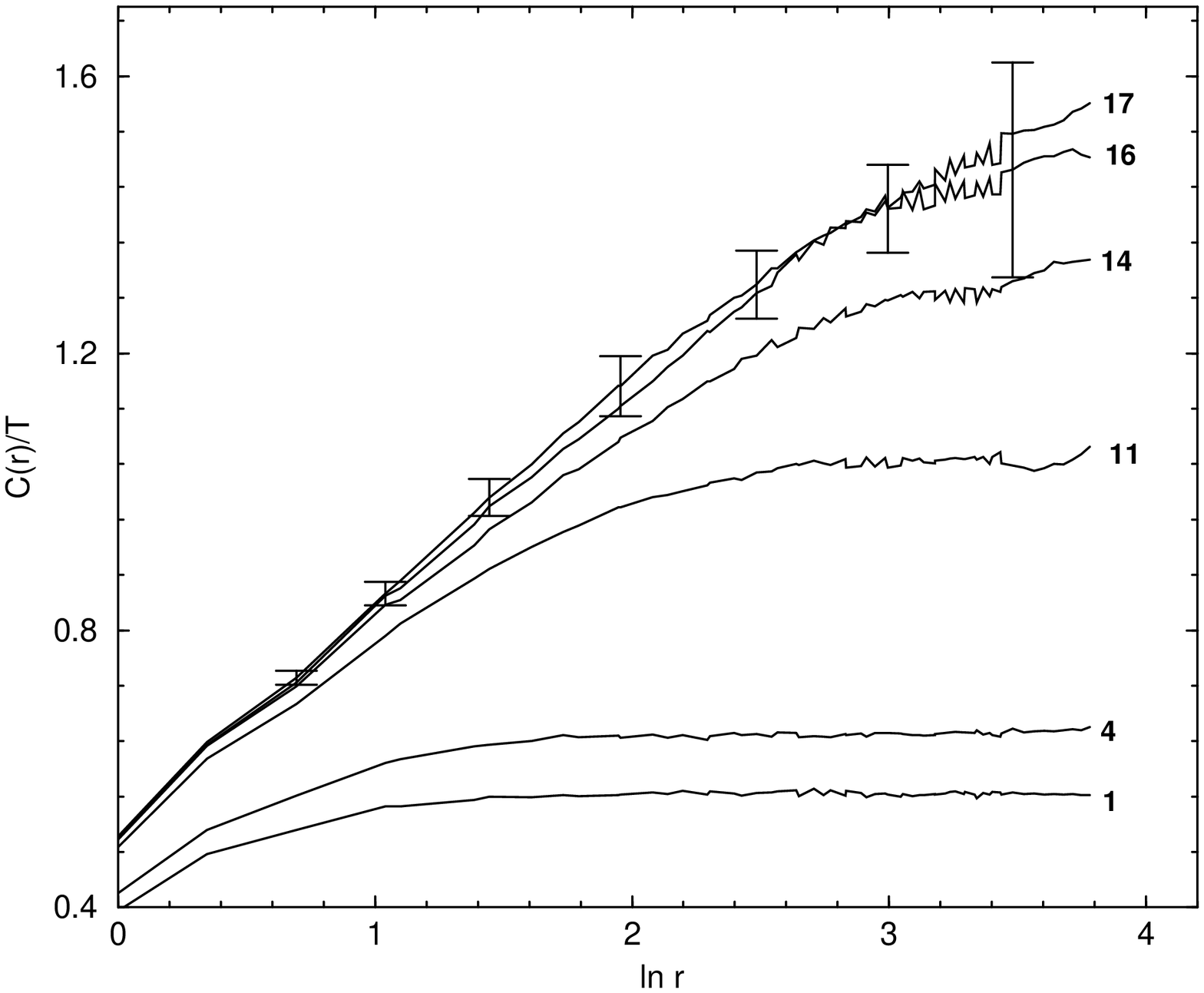, width=2.6in, angle=-90}
\epsfig{file=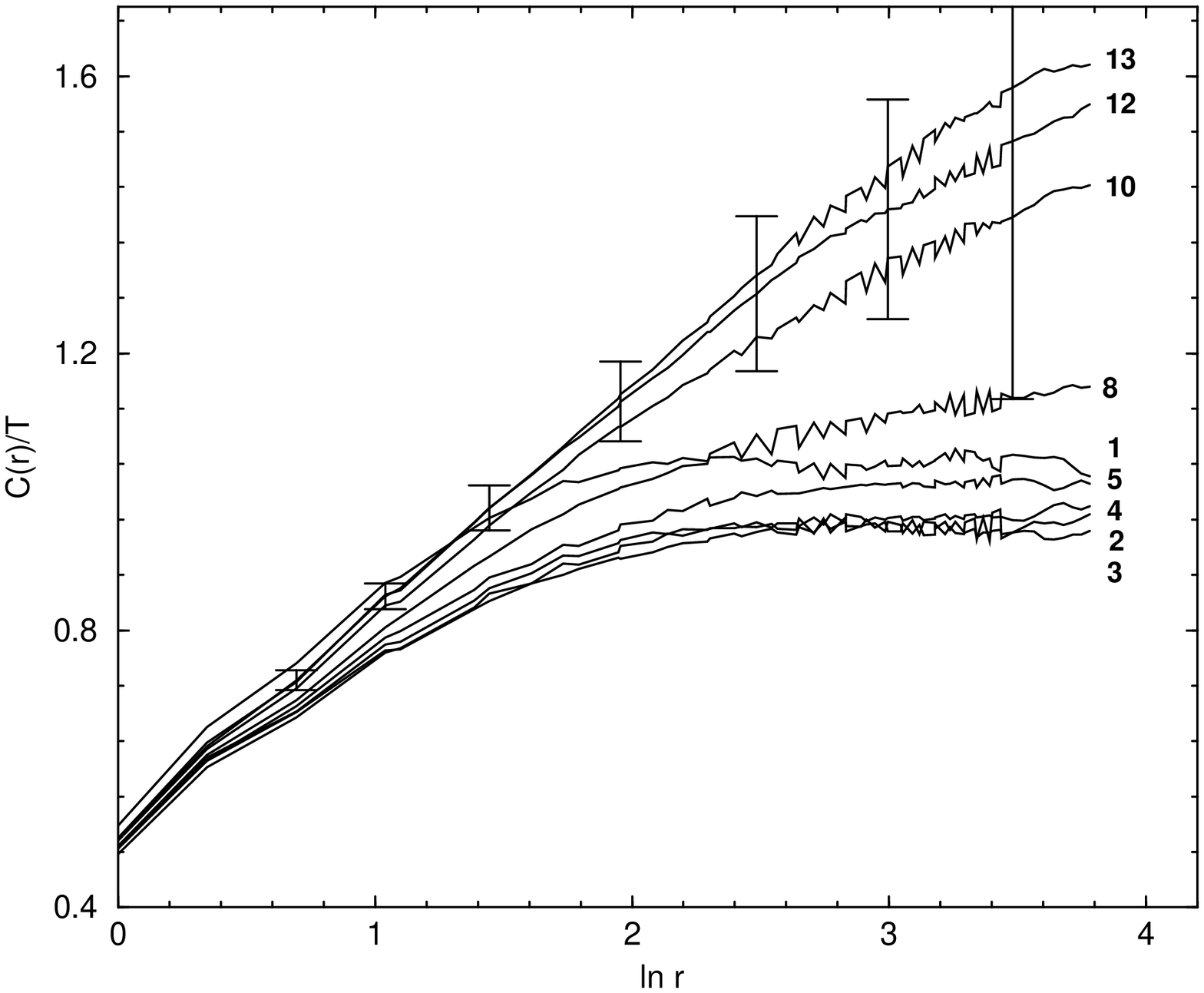, width=2.7in, angle=-90}
\caption{Height difference correlation functions (scaled by $T$) 
vs $\ln r$, $r$ being distance,
for the $\epsilon=0.2$ (upper panel) and $\epsilon=0.8$ (lower 
panel) RsGM.
Temperatures are indicated at the right 
side of the plots. The curve marked with temperature $16$ in the plot
for the RsGM is the correlation function for the OsGM at that temperature,
showing clearly that both overlap.}
\label{fig10}
\end{center}
\end{figure}

\begin{figure}
\begin{center}
\epsfig{file=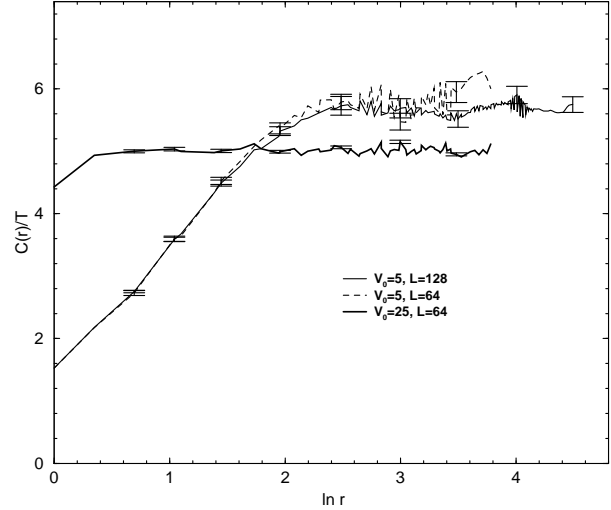, width=2.7in, angle=-90}
\caption{Height difference correlation functions (scaled by $T$) 
vs $\ln r$, $r$ being distance,
for different values of $V_0$ and $L$, as indicated in the plot. 
In all cases, $T=1$.}
\label{fig10bis}
\end{center}
\end{figure}

A second, novel finding arises when 
considering our numerical results for $V_0=5$, which strongly suggest the
possibility of two different low temperature phases. In our comments in 
the preceding paragraph, we took
$T^{V_0=5}_{SR}=8\pm 1$
interpreting that the superroughening transition implies a change
from a $\ln^2$ behavior of the height difference correlation function to
a $\ln$ form (and the rest of EW features).
However, the lack of size dependence
of the roughness and the specific heat on temperature below $T^*=5\pm 1$, 
with peaks absent for smaller values of $V_0$, raises the possible
existence of another phase transition. If one looks at the correlation 
functions in Fig.\ \ref{fig13}, it turns out that for temperatures 
below $T^*=4$ the correlation length is finite, in agreement with 
the roughness independence on the system size.
Whereas the 
range of correlations above $T^*$, which we believe is infinite, could 
be a subject of debate as we only have studied sizes up to $L=256$, our
claim of finite correlation lengths below $T^*$ is difficult 
to dispute. Further evidence in this regard is shown in Fig.\ \ref{fig10bis},
where curves for $V_0=5$ at $T=1$ are compared for two different system
sizes. At this point, it is interesting to recall that in a previous
paper \cite{us1} it was found that at $T=0$, the RsGM with $V_0=1$ 
exhibits a finite correlation length of about 20 lattice units 
(the reader may find Figs.\ 2 and 3 of \cite{us1} illustrative). 
Having this in mind, it is not unreasonable to conjecture that there
is a $T^*$ for the $V_0=1$ case, which could be below $T=1$ or close 
to 1 (the upper curve in Fig.\ \ref{fig4} might already show a 
finite correlation length). We stress that this phase has not 
been previously reported in works at $T=0$:
Thus, Zeng {\em et al.}\ \cite{Zeng} studied a discrete model but, 
being different from the Gaussian, their low temperature results can 
not be compared to ours, 
and the results of Rieger {\em et al.} 
\cite{Rieger1,Rieger2} do not allow to conclude anything in this 
respect. Intuitively, one can expect a finite 
correlation length phase at low temperatures and large $V_0$; in the
limit $V_0\to\infty$, the surface follows the disorder (i.e., 
$h(\rb)=h(\rb)^{(0)}+2n(\rb)\pi$ everywhere), but the gradient
term smoothens out the lack of correlations of $h^{(0)}(\rb)$,
the competition of these two effects yielding a finite correlation length.
In a loose 
sense, this could be interpreted as Anderson localization 
taking over the coupling between neighboring sites with increasing $V_0$. 
This picture is confirmed by simulations for $V_0=25$
(Fig.\ \ref{fig10bis}): For such a large value of $V_0$ the 
correlation length is only 1 lattice unit, i.e., correlations 
reach only nearest neighbors. 
Clearly, the data presented here is not
conclusive, but the conjecture that there are two transitions
whose critical temperatures depend on $V_0$ is not unreasonable 
and deserves further 
consideration. 

We have also found that 
the transition temperature and the correlation functions 
depend on the disorder strength.
This is not unexpected, in so far as the change in the disorder
distribution interacts with the periodicity of the sine potential, 
and therefore it is clear that when $\epsilon=2\pi$, i.e., in the
standard RsGM, we are in a special case. In this respect, our
results appear to indicate that the RsGM (with $\epsilon=2\pi$) is 
a very specific model, and that its behavior at low temperatures 
might not be representative of what one would find in an actual 
experiment, where the disorder can not be so precisely controlled. 
Another conclusion we may draw from our work is that
there might 
be two classes of behavior at low temperatures for small and large
$\epsilon$: Small $\epsilon$ models would behave very similarly to
the OsGM, whereas large values of $\epsilon$ would give rise to a 
more complex phenomenology with, e.g., nonmonotonic behavior of the
correlation functions. 

As a final conclusion, we remark that the most relevant result of 
the present work is the determination of the transition temperature
from the high temperature phase to the low temperature phase (or 
phases) of both the OsGM and the RsGM. This 
poses a number of questions to be addressed either with greater
numerical capabilities or with new analytical tools. 
We believe that the complex phase diagram of the RsGM is being partially
unveiled, as our research supports previous findings such as the 
non universality of the transition temperature. Significantly, once
we know where to look for the low temperature phase of the RsGM, we
can investigate the nature of
that phase (or phases). It is often stated
that the RsGM is ``glassy'' in this regime; however, this assertion 
has never been really proven nor fully detailed and, furthermore, 
if the RG picture is qualitatively correct, it has to be recalled that
it is a replica-symmetric theory, which implies that the
superrough phase would not be glassy in the replica sense \cite{jjfises}. 
We have obtained preliminary evidence that there are 
long-lived metastable states in the low temperature phase of the 
RsGM \cite{unpub}, but in view of our present results and their
non-universality we will examine this question more carefully in future work.
Investigation of the dynamics of the RsGM will also be important; we 
recall that Batrouni and Hwa \cite{Batrouni} did find evidence for a
superroughening transition in the dynamics of the model, and hence it would 
be worth checking their results for larger values of $V_0$. We hope
that those analyses, along with measurements of nonlinear susceptibilities
and of relaxation dynamics, will shed light on this difficult problem.
Work along these lines is in progress. 

\begin{acknowledgments}
We thank Rodolfo Cuerno, Juan Jes\'us Ruiz Lorenzo, and Ra\'ul Toral
for discussions.
Work at Legan\'es was supported by project PB96-0119.
Work at Los Alamos was done under the auspices of the U.S.\ Department
of Energy. A.S.\ and E.M.\ acknowledge the support by NATO CRG 971090,
by DGESIC (Spain) through a ``Acci\'on Integrada Hispano-Brit\'anica,''
and by Los Alamos National Laboratory for a stay at Los Alamos, where
part of this work was carried out.
\end{acknowledgments}


\end{multicols}
\newpage

\end{document}